\documentclass[conference]{IEEEtran}
\IEEEoverridecommandlockouts
\usepackage[letterpaper, left=0.62in, right=0.62in, bottom=1in, top=0.75in]{geometry}
\usepackage{amssymb,amsmath}
\usepackage{graphicx}
\usepackage{cite}
\usepackage{txfonts}
\usepackage{mathrsfs}
\usepackage{fontenc}
\usepackage[binary-units]{siunitx}
\usepackage[caption=false,font=footnotesize]{subfig}
\usepackage{tabularx}
\usepackage{multirow}
\usepackage{diagbox}
\usepackage{algorithm2e}
\usepackage{bbm}
\usepackage{xcolor}

\begin{document}
\title{Deep Learning-Aided Delay-Tolerant Zero-Forcing Precoding in Cell-Free Massive MIMO}
\author{\IEEEauthorblockN{Wei Jiang\IEEEauthorrefmark{1}\IEEEauthorrefmark{2} and Hans D. Schotten\IEEEauthorrefmark{2}\IEEEauthorrefmark{1}}
\IEEEauthorblockA{\IEEEauthorrefmark{1}German Research Center for Artificial Intelligence (DFKI), Trippstadter Str. 122,  Kaiserslautern, 67663 Germany\\
  }
\IEEEauthorblockA{\IEEEauthorrefmark{2}Technische Universit\"at (TU) Kaiserslautern, Building 11, Paul-Ehrlich Street, Kaiserslautern, 67663 Germany\\
 }
\thanks{This work of the authors was supported by the German Federal Ministry of Education and Research (BMBF) through the \emph{KICK} project (Grant no. \emph{16KIS1105}).
}
}

\maketitle

\begin{abstract}
In the context of cell-free massive multi-input multi-output (CFmMIMO), zero-forcing precoding (ZFP) is superior in terms of spectral efficiency. However, it suffers from channel aging owing to fronthaul and processing delays. In this paper,  we propose a robust scheme coined delay-tolerant zero-forcing precoding (DT-ZFP), which exploits deep learning-aided channel prediction to alleviate the effect of outdated channel state information (CSI). A predictor consisting of a bank of user-specific predictive modules is specifically designed for such a multi-user scenario. Leveraging the degree of freedom brought by the prediction horizon, the delivery of CSI and precoded data through a fronthaul network and the transmission of user data and pilots over an air interface can be parallelized. Therefore,  DT-ZFP not only effectively combats channel aging but also avoids the inefficient ``\textit{Stop-and-Wait}" mechanism of the canonical ZFP in CFmMIMO.
\end{abstract}

\section{Introduction}

Cell-free massive multi-input multi-output (CFmMIMO) \cite{Ref_ngo2017cellfree, Ref_jiang2021cellfree, Ref_nayebi2017precoding} has gained much attention recently  due to its potential of becoming a technical enabler for the six-generation (6G) system \cite{Ref_jiang2021road}. It employs a large number of distributed access points (APs) to simultaneously serve a few users in a geographical area over the same time-frequency resource. The dominant number of APs over user equipment (UE) makes linear precoding, i.e., conjugate beamforming (CBF) and zero-forcing precoding (ZFP), perform nearly as good as dirty-paper coding \cite{Ref_costa1983writing} in CFmMIMO. It is extensively verified that ZFP is superior to CBF with much higher spectral efficiency \cite{Ref_nayebi2017precoding}.
However, all APs in ZFP are required to send their local channel state information (CSI) to a central processing unit (CPU) via a fronthaul network and then stop-and-wait until the CPU sends back precoded data. This particular process induces a considerable delay, raising channel aging in fast-fading environments. This problem will be more challenging in 6G, where \textit{high mobility}, e.g., high-speed trains and unmanned aerial vehicles, and \textit{high frequency}, such as millimeter-wave and terahertz signals \cite{Ref_jiang2022initial_ICC}, further aggravate the fading of wireless channels. Thus, the canonical ZFP in CFmMIMO suffers from two major problems: (1) the performance degradation due to channel aging, and (2) the inefficient time resource utilization because of the ``\textit{Stop-and-Wait}" mechanism.

In the literature, the effect of channel aging on co-located massive MIMO, see \cite{Ref_Papazafeiropoulos2017impact, Ref_chopra2018performance}, and the uplink of CFmMIMO have been reported \cite{Ref_zheng2020cellfree}. To the best knowledge of the authors, the impact of aged CSI on the downlink of CFmMIMO has not been discussed until the authors of this paper provided the first work in \cite{Ref_jiang2021impactcellfree}. The theoretical analysis and numerical evaluation in \cite{Ref_jiang2021impactcellfree} revealed that the fronthaul and processing delays deteriorate performance substantially. Consequently, the effect of channel aging should be seriously considered and an effective mitigation method is mandatory in the practical deployment of CFmMIMO.

As a follow-up of \cite{Ref_jiang2021impactcellfree}, the aim of this paper is to propose a robust transmission scheme against channel aging, coined delay-tolerant zero-forcing precoding (DT-ZFP). We exploit deep learning (DL)-aided channel prediction \cite{Ref_jiang2020deep} to improve the quality of CSI. Although prior works investigated modeled-based \cite{Ref_jiang2019comparison} and data-driven prediction \cite{Ref_jiang2019neural}, these methods only focus on a single-user setup where the difference of distance-dependent large-scale fading among different users is not considered. In this paper, we extend the single-user prediction to meet the requirements of DT-ZFP by specifically designing a multi-user predictor consisting of a bank of user-specific DL predictive modules. Leveraging the degree of freedom brought by the prediction horizon, the delivery of precoded data and CSI through a fronthaul network and the transmission of user data and pilot signals over air interface can be parallelized. As a result, DT-ZFP can not only effectively combat channel aging but also avoid the  inefficient resource usage due to ``\textit{Stop-and-Wait}" mechanism. The superiority of the proposed scheme is justified by simulations.

The remainder of this paper is structured as follows: Section II introduces the system model. Section III presents the communications process of DT-ZFP and the design of the multi-user deep-learning predictor. Simulation setup and numerical results are demonstrated in Section IV. Finally, the conclusions are drawn in Section V.
\begin{figure}[!t]
    \centering
    \includegraphics[width=0.45\textwidth]{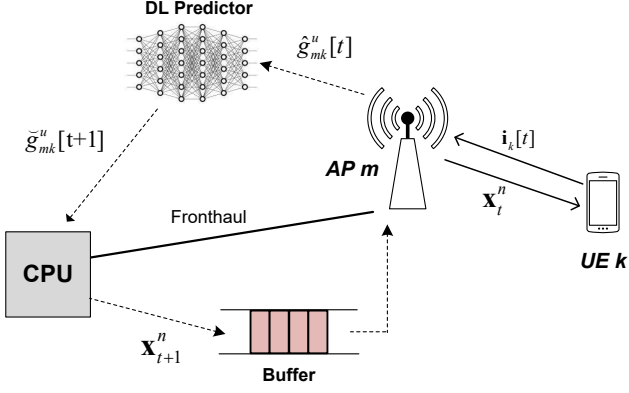}
    \caption{Schematic diagram of a cell-free massive MIMO system consisting of a CPU, $M$ APs, and $K$ users. The proposed  scheme employs deep learning to convert estimated CSI to predicted CSI, and buffers the transmitted symbols precoded based on the predicted CSI. }
    \label{fig:SystemModel}
\end{figure}

\section{System Model}

We consider a CFmMIMO system where a large number of $M$ single-antenna APs and $K$ single-antenna UEs are randomly distributed over a geographical area, with $M\gg K$.  The signal transmission and reception of the APs are coordinated by a CPU via a fronthaul network to simultaneously serve the users over the same time-frequency resource, as shown in \figurename \ref{fig:SystemModel}. The downlink transmission from the APs to the UEs and the uplink transmission from  the UEs to the APs are separated by time-division duplex (TDD) operation. Channel reciprocity is exploited in TDD to avoid a remarkably high overhead of downlink pilots that are proportional to the number of APs. A radio frame is divided into three phases: uplink training, uplink transmission, and downlink transmission. In the uplink training, the UEs send orthogonal pilot sequences to the APs, so that each AP can estimate the instantaneous CSI, which is employed to precode the information symbols in the downlink and detect the received signals in the uplink. As \cite{Ref_jiang2021impactcellfree}, this paper focuses on the downlink transmission where the proposed DT-ZFP is applied, whereas the uplink transmission is skipped since the fronthaul delay does not affect the signal detection at the CPU.

Without losing generality, we assume that small-scale fading follows frequency-flat block fading. A frequency-selective channel can be transformed into a magnitude of flat-fading sub-channels through orthogonal frequency-division multiplexing \cite{Ref_jiang2016ofdm}, making this assumption reasonable.
We write $g_{mk}=\sqrt{\beta_{mk}} h_{mk}$ to denote the channel coefficient between AP $m$, $\forall m=1,2,\ldots,M$ and UE $k$, $\forall k=1,2,\ldots,K$, where $\beta_{mk}$ and $h_{mk}$ represent large-scale and small-scale fading, respectively. Usually,
$h_{mk}$ is modelled as a circularly symmetric complex Gaussian random variable with zero mean and unit variance, i.e., $h_{mk} \sim \mathcal{CN}(0, 1)$. Large-scale fading equals $\beta_{mk}=10^\frac{\mathcal{P}_{mk}+\mathcal{S}_{mk}}{10}$, where $\mathcal{S}_{mk}$ denotes shadowing fading with zero mean and variance $\sigma_{sd}^2$, namely $\mathcal{S}_{mk}\sim \mathcal{N}(0,\sigma_{sd}^2)$,  and $\mathcal{P}_{mk}$ represents path loss, which can be computed by the COST-Hata model \cite{Ref_ngo2017cellfree} as
\begin{equation} \label{eqn:CostHataModel}
    \mathcal{P}_{mk}=
\begin{cases}
-\mathcal{P}_0-35\lg(d_{mk}) &  d_{mk}>d_1 \\
-\mathcal{P}_0-15\lg(d_1)-20\lg(d_{mk}) &  d_0<d_{mk}\leq d_1 \\
-\mathcal{P}_0-15\lg(d_1)-20\lg(d_0) &  d_{mk}\leq d_0
\end{cases},
\end{equation}
where $\lg$ stands for the common logarithm, $d_{mk}$ is the propagation distance,  $d_0$ and $d_1$ denote the break points, and the reference path loss at \SI{1}{\meter} is given by
\begin{IEEEeqnarray}{ll}
 \mathcal{P}_0=46.3&+33.9\lg\left(f_c\right)-13.82\lg\left(h_{AP}\right)\\ \nonumber
 &-\left[1.1\lg(f_c)-0.7\right]h_{UE}+1.56\lg\left(f_c\right)-0.8
\end{IEEEeqnarray} with carrier frequency $f_c$, the AP antenna height $h_{AP}$, and the UE antenna height $h_{UE}$.

\section{Delay-Tolerant Zero-Forcing Precoding}
Exploiting the potential of DL-based channel prediction, we propose a delay-tolerant transmission scheme for the downlink of CFmMIMO. This section first presents the principle of DT-ZFP through its communications process and then introduces the multi-user predictor that is built by a bank of user-specific DL predictive modules.

\begin{figure}[!b]
    \centering
    \includegraphics[width=0.45\textwidth]{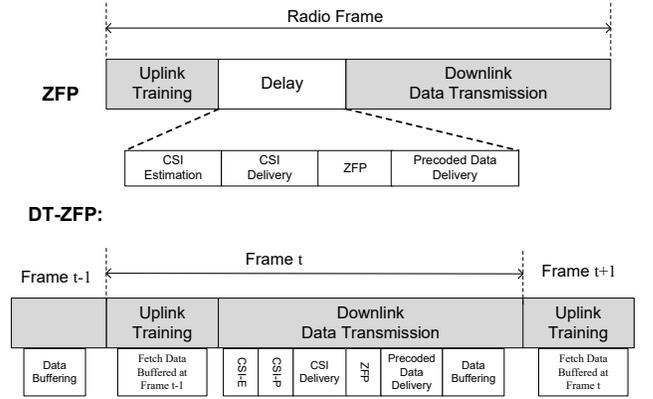}
    \caption{The frame structure of DT-ZFP in comparison with that of the conventional ZFP.  Due to the prediction horizon enabled by channel prediction, the delivery of precoded data and CSI through the fronthaul network and the transmission of user data and pilots over air interface can be parallelized, avoiding the inefficient \textit{Stop-and-Wait} operation in ZFP.}
    \label{fig:FrameStructure}
\end{figure}

\subsection{The Communications Process}
As prior works such as \cite{Ref_ngo2017cellfree, Ref_jiang2021cellfree, Ref_nayebi2017precoding}, we assume that $\beta_{mk}$ is perfectly known, and the fronthaul network is error-free and capacity-infinite so as to particularly focus on the fronthaul delay without the disturb of practical constraints \cite{Ref_masoumi2020performance}.  We only consider the uplink training and downlink data transmission hereinafter, whereas the uplink transmission is neglected as explained before.  As illustrated in \figurename \ref{fig:FrameStructure}, the DT-ZFP scheme operates as follows:
\begin{enumerate}
\item \textit{Uplink Training:} The communications process is organized in radio frames. At the training phase of radio frame $t$, the UEs  transmit orthogonal pilot sequences $\mathbf{i}_k[t]$, $k=1,\ldots,K$ simultaneously towards the APs. Due to the orthogonality, we have $\mathbf{i}_k^H\mathbf{i}_{k'}=0$, $\forall k'\neq k$.  A typical AP $m$ observes
\begin{equation}
    \mathbf{y}_{m}^u[t] = \sqrt{p_u} \sum_{k=1}^K g_{mk}^u[t] \mathbf{i}_k[t]+\mathbf{n}_{m}[t],
\end{equation}
where $p_u$ is the UE power constraint, $g_{mk}^u[t]$ denotes the instantaneous channel gain between AP $m$ and UE $k$ during the uplink training at frame $t$, and additive white Gaussian noise (AWGN) has zero mean and variance $\sigma_n^2$, i.e., $\mathbf{n}_m\in \mathcal{CN}\left(\mathbf{0}, \sigma_n^2\mathbf{I}\right)$.
\item \textit{CSI Estimation (CSI-E):} The $m^{th}$ AP can compute the linear minimum mean-square error (MMSE) estimate of $g_{mk}^u[t]$, $\forall k$ as \cite{Ref_nayebi2017precoding}
\begin{equation}
    \hat{g}_{mk}^u[t] = \left(\frac{\sqrt{p_u}\beta_{mk}}{p_u \beta_{mk} + \sigma_n^2}\right)\mathbf{i}_k^H[t] \mathbf{y}_{m}^u[t].
\end{equation} Denoting the channel estimation error as $\tilde{g}_{mk}^u = g_{mk}^u - \hat{g}_{mk}^u$, we have $\tilde{g}_{mk}^u\in \mathcal{CN}\left(0,\beta_{mk}-\alpha_{mk}\right)$, where $\beta_{mk}$ and $\alpha_{mk}=\frac{p_u\beta_{mk}^2}{p_u \beta_{mk} + \sigma_n^2}$ are the variances of $g_{mk}^u$ and $\hat{g}_{mk}^u$, respectively.
\item \textit{CSI Prediction (CSI-P):} Unlike ZFP that sends the estimated CSI directly to the CPU, DT-ZFP conducts channel prediction before the delivery of CSI. Each AP feeds its local CSI, e.g.,  $\left\{ \hat{g}_{m1}^u[t],\hat{g}_{m2}^u[t],\ldots,\hat{g}_{mK}^u[t]  \right\}$ for AP $m$, into a local predictor to get prediction values
\begin{equation}
    \check{g}_{mk}^u[t+1]=f_m\Bigl(\hat{g}_{mk}^u[t]\Bigr), \:\:\forall k,
\end{equation} where $f_m(\cdot)$ represents the input-output function of the multi-user channel predictor running at AP $m$, which will be elaborated in the next sub-section.
\item \textit{CSI Delivery:} Each AP sends its predicted CSI, e.g.,
\begin{equation}
    \check{\mathbf{g}}_{m}[t+1]=\left[\check{g}_{m1}^u[t+1],\ldots,\check{g}_{mK}^u[t+1]\right]^T\in \mathbb{C}^{K\times 1}
\end{equation} at AP $m$, to the CPU through the fronthaul network. Thus, the CPU gets the global CSI prediction
\begin{equation}
    \check{\mathbf{G}}_{t+1}=\biggl[\check{\mathbf{g}}_{1}[t+1],\ldots,\check{\mathbf{g}}_{M}[t+1]\biggr] \in \mathbb{C}^{K\times M}.
\end{equation}
\item \textit{Zero-Forcing Precoding:} Assume the downlink transmission has $N$ symbol periods, we denote the symbol vector at period $n$, $n=1,\ldots,N$ as
\begin{equation}
    \mathbf{s}_{t+1}^n=\Bigl[s_1^n[t+1],\ldots,s_K^n[t+1]\Bigr]^T,
\end{equation}
where $s_k^n[t+1]$ is the information symbol intended for user $k$ at the $n^{th}$ symbol period of frame $t+1$, satisfying $\mathbb{E}[\vert s_k \vert^2]=1$. The CPU precodes $\mathbf{s}_{t+1}^n$ to get
\begin{equation}
    \mathbf{x}_{t+1}^n=\Bigl[x_1^n[t+1],\ldots,x_M^n[t+1]\Bigr]^T
\end{equation}
through
\begin{equation}
    \mathbf{x}_{t+1}^n=\check{\mathbf{G}}^H_{t+1}\left(\check{\mathbf{G}}_{t+1}\check{\mathbf{G}}^H_{t+1}\right)^{-1}\boldsymbol{\Psi}_{t+1}\mathbf{s}_{t+1}^n,
\end{equation}
where $x_m^n[t+1]$ denotes the precoded symbol to be transmitted by AP $m$ at the $n^{th}$ symbol period of frame $t+1$, $\boldsymbol{\Psi} \in \mathbb{C}^{K\times K}$ is a diagonal matrix consisting of power-control coefficients, i.e.,
\begin{equation}
    \boldsymbol{\Psi}_{t+1}=\mathrm{diag}\Bigl\{\psi_1[t+1],\psi_2[t+2],\ldots,\psi_K[t+1]\Bigr\}.
\end{equation}
\item \textit{Precoded Symbol Buffering:} The CPU distributes the precoded symbols towards their corresponding APs, namely
\begin{equation}
    \left\{x_m^1[t+1],x_m^2[t+1],\ldots,x_m^N[t+1] \right\}
\end{equation} for AP $m$, through the fronthaul network. AP $m$ receives $x_m^n[t+1]$, $n=1,\ldots,N$ from the CPU and stores these symbols in its buffer. Note that these precoded symbols will be transmitted at the next frame indexed $t+1$, while each AP transmits the precoded symbols $x_m^n[t]$, $n=1,\ldots,N$ buffered at the previous frame $t-1$.
\item As shown in \figurename \ref{fig:FrameStructure},  the APs in ZFP need to \textit{stop and wait} for the arrival of precoded symbols from the CPU after the delivery of CSI. Due to the feedback and processing delays, there is a time gap between the completion of receiving pilot sequences (uplink training) and the start of transmitting the precoded symbols (downlink transmission). More details of modeling this gap can refer to Fig. 1 of our previous work \cite{Ref_jiang2021impactcellfree}. However, DT-ZFP can start the downlink transmission immediately once the uplink training is completed since the transmitted symbols for the current frame, i.e., $x_m^n[t]$, $n=1,\ldots,N$, are already buffered at the previous frame $t-1$.
As a particular degree of freedom enabled by the prediction horizon, the downlink transmission of DT-ZFP can be performed in parallel with other processing, i.e., the CSI estimation and delivery, and the precoding and distributing of symbols, as depicted from step 2) to step 6). Consequently, the inefficient \textit{stop-and-wait} operation in ZFP is avoided, resulting in more efficient usage of time resource.
\end{enumerate}

\SetKwComment{Comment}{/* }{ */}
\RestyleAlgo{ruled}
\begin{algorithm}
\caption{Delay-Tolerant Zero-Forcing Precoding} \label{alg:IRS001}
\SetKwInOut{Input}{input}
\Input{M, K, and $\beta_{mk}$, $\forall m=1,\ldots,M,\: k=1,\ldots,K$}
\For{Radio frame $t$}{
\ForEach{UE $k=1,2,\ldots,K$}
 {Send pilot sequence $\mathbf{i}_k$\;
  }
\ForEach{AP $m=1,2,\ldots,M$}{
Start transmission after uplink-downlink switch\;
Fetch $x_m^n[t]$, $n=1,\ldots,N$ from the buffer\;
Transmit $x_m^n[t]$ with power $p_d$\;
Estimate local CSI $\hat{g}_{mk}^u[t]$, $\forall k=1,\ldots,K$\;
Predict $\check{g}_{mk}^u[t+1]=f_m\Bigl(\hat{g}_{mk}^u[t]\Bigr)$\;
Deliver $\check{\mathbf{g}}_{m}[t+1]$ to the CPU\;}
CPU get $\check{\mathbf{G}}_{t+1}=\biggl[\check{\mathbf{g}}_{1}[t+1],\ldots,\check{\mathbf{g}}_{M}[t+1]\biggr]$\;
ZF Precoding: $\mathbf{x}_{t+1}^n=\check{\mathbf{G}}^H_{t+1}\left(\check{\mathbf{G}}_{t+1}\check{\mathbf{G}}^H_{t+1}\right)^{-1}\boldsymbol{\Psi}_{t+1}\mathbf{s}_{t+1}^n$\;
CPU distribute $\left\{x_m^1[t+1],\ldots,x_m^N[t+1]\right\}$ to AP $m$\;
AP $m$ buffer $\left\{x_m^1[t+1],\ldots,x_m^N[t+1]\right\}$\;
}
\end{algorithm}

The proposed DT-ZFP scheme is summarized as \algorithmcfname \ref{alg:IRS001}. As a result, a typical user $k$ obtains the received symbol at the $n^{th}$ period of frame $t$ as
\begin{align} \label{eqn:RxSignal_matrix} \nonumber
r_k^n[t] &= \sqrt{\rho_d}\mathbf{g}_k^T[t] \mathbf{x}_{t}^n + n_k\\
&= \sqrt{\rho_d}\mathbf{g}_k^T[t] \check{\mathbf{G}}_t^H\left(\check{\mathbf{G}}_t\check{\mathbf{G}}_t^H\right)^{-1}\boldsymbol{\Psi}_t\mathbf{s}_t^n+n_k,
\end{align}
where $\mathbf{g}_k[t]=\Bigl[g_{1k}[t],g_{2k}[t]\ldots,g_{Mk}[t]\Bigr]^T\in \mathbb{C}^{M\times 1}$ denotes the channel fingerprint of user $k$ at frame $t$, $p_d$ is the AP power constraint, and $n_k$ is AWGN with zero mean and variance $\sigma_n^2$, i.e., $n_k\in \mathcal{CN}\left(0, \sigma_n^2\right)$.

\subsection{DL-based Multi-User CSI Predictor}
\begin{figure}[!tbph]
    \centering
    \includegraphics[width=0.45\textwidth]{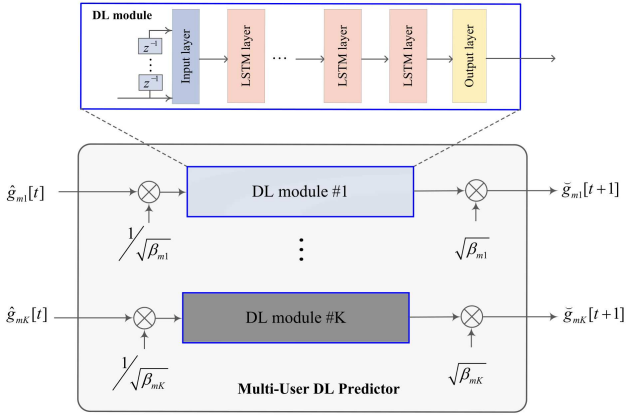}
    \caption{The structure of a multi-user predictor operating at a typical AP $m$, which is mainly comprised of $K$ independent deep-learning modules. Each module consists of an input layer, an output layer, and multiple LSTM or GRU hidden layers.}
    \label{fig:CSIpredictor}
\end{figure}

We exploit deep learning-aided channel prediction \cite{Ref_jiang2020deep} to improve the quality of CSI. Although prior works investigated modeled-based \cite{Ref_jiang2019comparison} and data-driven prediction \cite{Ref_jiang2019neural}, these methods focus on  single-user scenarios where only small-scale fading is considered with the assumption that channel gains follow standard complex Gaussian distribution, i.e., $h\sim \mathcal{CN}(0,1)$. However, a practical wireless system needs to accommodate a lot of users and simultaneously serve multiple active users. Due to the near-far effect, distance-dependent large-scale fading among multiple users might differ by several orders of magnitude or tens of decibels (dB). In CFmMIMO, different AP-UE pairs have different propagation distances, where previous single-user channel prediction cannot be directly applied. We therefore design a multi-user predictor consisting of a bank of user-specific DL predictive modules to deal with such power-gain difference.

\figurename \ref{fig:CSIpredictor} shows the structure of the multi-user predictor operating at a typical AP $m$. The estimated CSI $\hat{g}_{mk}[t]$, $\forall k$ is first normalized by multiplexing a factor of $1/\sqrt{\beta_{mk}}$. Recalling that $\hat{g}_{mk}[t]=\sqrt{\beta_{mk}}\hat{h}_{mk}[t]$, the input data for the $k^{th}$ DL module equals $\hat{h}_{mk}[t]$, following  $\mathcal{CN}(0,1)$. Thus, each DL module, which generally consists of an input layer, $L$ hidden layers, and an output layer, can perform single-user prediction independently.
Feeding  $\hat{h}_{mk}[t]$ into the input feed-forward layer obtains $\mathcal{I}\left(\hat{h}_{mk}[t]\right)=\delta_h \left( \mathbf{w}^{i} \hat{h}_{mk}[t] +\mathbf{b}^{i} \right)$, where $\mathbf{w}^{i}$ and  $\mathbf{b}^{i}$ denote the vectors of weights and biases of the input layer, and $\delta_h$ stands for an activation function. The first hidden layer generates $\mathcal{L}^{(1)}\left(\mathcal{I}\left(\hat{h}_{mk}[t]\right)\right)$ as the response to the activation of the input layer. The structure of a hidden layer built by long short-term memory (LSTM) or gated recurrent unit (GRU), and the detail definition of  $\mathcal{L}\left(\cdot\right)$ can refer to the previous work of the authors \cite{Ref_jiang2021simple}. The activation goes through the network until the output layer modelled by $\mathcal{T}(\cdot)$ gets a predicted value
\begin{equation}
    \check{h}_{mk}[t+1]=\mathcal{T}\left(\mathcal{L}^{(L)}\left( \ldots \mathcal{L}^{(2)}\left(\mathcal{L}^{(1)}\left(\mathcal{I}\left(\hat{h}_{mk}[t]\right)\right)\right)\right)\right).
\end{equation}
The predicted CSI is obtained by multiplexing a factor of $\sqrt{\beta_{mk}}$, i.e.,
\begin{equation}
    \check{g}_{mk}[t+1] = \sqrt{\beta_{mk}} \check{h}_{mk}[t+1].
\end{equation}

As revealed in \cite{Ref_jiang2021simple}, the computational complexity of a DL module is low since CSI prediction needs only a few (generally two or three) hidden layers with a small number of neurons, unlike other high-complexity applications such as face recognition or natural language processing that requires tens of hidden layers with a large number of neurons per layer.

\section{Performance Evaluation and Comparison}

The performance of the proposed scheme is evaluated through Mont-Carlo simulations. This section introduces the simulation setup and illustrates some representative numerical results to observe the gain of DT-ZFP on the downlink of a CFmMIMO system.
Consider a square urban area of $\SI{1}{\kilo\meter}\times \SI{1}{\kilo\meter}$ where $M=128$ distributed APs serve $K=16$ UEs at the same time-frequency resource. To calculate large-scale fading using (\ref{eqn:CostHataModel}), we take values $d_0=10\mathrm{m}$, $d_1=50\mathrm{m}$, and $\mathcal{P}_0=140.72\mathrm{dB}$ with $f_c=1.9\mathrm{GHz}$, $h_{AP}=15\mathrm{m}$, and $h_{UE}=1.65\mathrm{m}$, while the standard derivation for shadowing fading is $\sigma_{sd}=8\mathrm{dB}$.
The power constraints of AP and UE are $p_d=0.2\mathrm{W}$ and $p_u=0.1\mathrm{W}$, respectively. Since the optimal max-min power control has high complexity,  a sub-optimal, low-complexity scheme \cite{Ref_nayebi2017precoding} is applied for power control, i.e., $\psi_{k}=\left( \max_m  \sum_{k=1}^{K} \delta_{km} \right)^{-1}$, $\forall k$, where $\boldsymbol \delta_m= \left[\delta_{1m},\ldots,\delta_{Km}\right]^T=\mathrm{diag}\left(\mathbb{E}\left[  \left(\check{\mathbf{G}}\check{\mathbf{G}}^H\right)^{-1}    \check{\mathbf{g}}_m \check{\mathbf{g}}_m^H   \left(\check{\mathbf{G}}\check{\mathbf{G}}^H\right)^{-1} \right]\right)$ and $\check{\mathbf{g}}_m$ stands for the $m^{th}$ column of $\check{\mathbf{G}}$.  The variance of AWGN is computed by $\sigma_n^2=\kappa\cdot B\cdot T_0\cdot N_f$ with the Boltzmann constant $\kappa$, signal bandwidth $B=20\mathrm{MHz}$, temperature $T_0=290 \mathrm{Kelvin}$, and  noise figure  $N_f=9\mathrm{dB}$.

To emulate a fast-fading scenario, the maximal Doppler shift is selected to $f_d{=}100 \mathrm{Hz}$, corresponding to a velocity of about 50 \si[per-mode=symbol]{\kilo\meter\per\hour} at carrier frequency of \SI{1.9}{\giga\hertz}. To get high prediction accuracy, the hyper-parameters of deep learning, mainly including the number of layers, the number of neurons per layer, activation functions, training algorithms, and the volume of training data, need to be carefully tuned. Generally, a training of a deep network is started from an initial state where all weights and biases are random. The prediction is compared with its desired value and the resultant error is propagated back through the network to update the weights by means of a training algorithm such as the Adam optimizer \cite{Ref_kingma2017adam}. Compared the prediction accuracy of different hyper-parameters, we select a 2-hidden-layer LSTM network with $25$ neurons at either layer and a training length of $5,000$. More details of  the dataset building and training process can refer to \cite{Ref_jiang2021predictive}.

\begin{figure}[!t]
    \centering
    \includegraphics[width=0.44\textwidth]{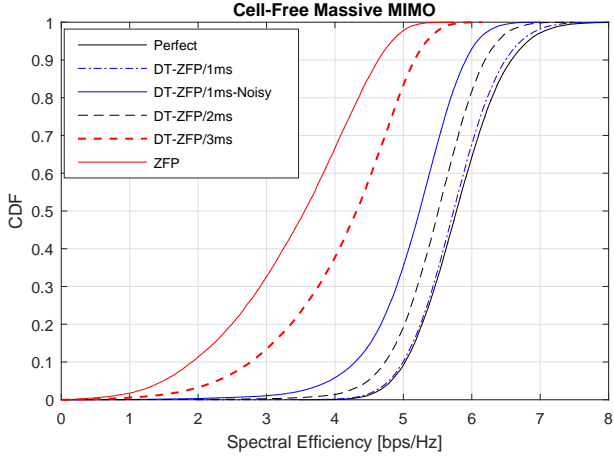}
    \caption{Performance evaluation of the proposed scheme under different prediction horizons of \SI{1}{\milli\second}, \SI{2}{\milli\second}, and \SI{3}{\milli\second}, in comparison with the benchmarks using the perfect CSI and outdated CSI. The performance is measured by the CDF of per-user spectral efficiency.}
    \label{fig:Results}
\end{figure}

\figurename \ref{fig:Results} provides the comparisons with respect to cumulative distribution functions (CDFs) of per-user spectral efficiency. The performance curves of ZFP using the perfect and outdated CSI are employed as the benchmarks to indicate the upper and lower boundaries, respectively. In the perfect case, the $5\%$-likely spectral efficiency, which is usually used to measure the cell-edge performance, and the $50\%$-likely or median spectral efficiency are around $4.8\mathrm{bps/Hz}$ and $5.8\mathrm{bps/Hz}$, respectively. For the ease of comparison, we set the overall delay in the conventional ZFP to \SI{1}{\milli\second}. Due to the outdated CSI, the $5\%$-likely and median spectral efficiencies decrease to $1.49\mathrm{bps/Hz}$ and $3.56\mathrm{bps/Hz}$,  amounting to a rate loss of approximately $70\%$ and $40\%$, respectively. The results further reveal the substantial impact of channel aging on the performance of ZFP in CFmMIMO.

We first observe the result of the proposed scheme under a prediction horizon of \SI{1}{\milli\second}. That is to say, the DL predictor predicts the upcoming CSI of \SI{1}{\milli\second} later based on the current estimated CSI. The signal-to-noise ratio (SNR) of the received pilot signals is set to be \SI{30}{\decibel}, which is reasonable in a practical wireless system with good conditions. It achieves the near-optimal performance with a $5\%$-likely and median spectral efficiency of  $4.78\mathrm{bps/Hz}$ and $5.73\mathrm{bps/Hz}$, respectively. The quality of estimated CSI affects the training of deep learning and the resultant prediction accuracy. Decreasing the SNR of the received pilot signals to \SI{15}{\decibel}, as indicated by \textit{Noisy} in the figure, its achievable spectral efficiency degrades to approximately $3.9\mathrm{bps/Hz}$ and $5.2\mathrm{bps/Hz}$. With the increase of prediction horizon, the prediction accuracy of the deep-learning predictor decreases. We further set a prediction horizon of \SI{2}{\milli\second} with a SNR of received pilot signals of \SI{30}{\decibel}, resulting in a $5\%$-likely and median spectral efficiency of  $4.48\mathrm{bps/Hz}$ and $5.5\mathrm{bps/Hz}$, respectively. Under a prediction horizon of \SI{3}{\milli\second}, which is long enough considering a channel coherence time of around \SI{10}{\milli\second} under the Doppler shift of $100 \si{\hertz}$, a $5\%$-likely and median spectral efficiency of  $2.28\mathrm{bps/Hz}$ and $4.32\mathrm{bps/Hz}$ are achieved. It can be concluded from the simulation results that the proposed scheme can effectively alleviate the impact of channel aging in a CFmMIMO system with a substantial performance gain.

\section{Conclusions}
This paper proposed a robust scheme called delay-tolerant zero-forcing precoding for the downlink transmission of cell-free massive MIMO systems. Exploiting deep learning-aided single-user channel prediction, we designed a multi-user predictor that is comprised of a bank of user-specific predictive modules, where the difference of distance-independent large-scale fading among users are settled. Leveraging the degree of freedom brought by the prediction horizon, the delivery of precoded data and CSI through a fronthaul network and the transmission of user data and pilots over air interface can be parallelized, avoiding the inefficient ``\textit{Stop-and-Wait}" mechanism of the conventional ZFP in CFmMIMO. Numerical results justified the effectiveness of the multi-user deep-leaning channel predictor, and the performance superiority of the proposed scheme in terms of achievable spectral efficiency.

\bibliographystyle{IEEEtran}
\bibliography{IEEEabrv,Ref_VTC2022Fall}

\end{document}